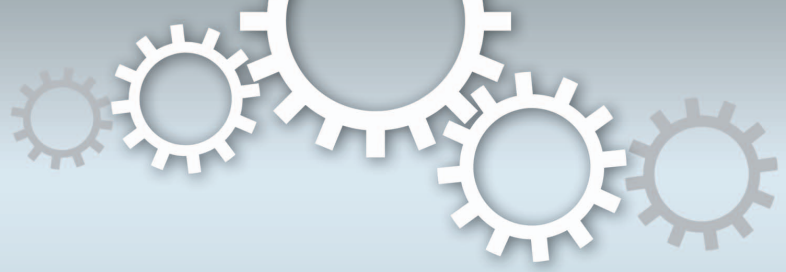

# SCIENTIFIC REPORTS

# Visualisation of edge effects in side-gated graphene nanodevices

Vishal Panchal[1,2], Arseniy Lartsev[3], Alessandra Manzin[4], Rositza Yakimova[5], Alexander Tzalenchuk[1,2] & Olga Kazakova[1]

[1]National Physical Laboratory, Teddington, TW11 0LW, United Kingdom, [2]Royal Holloway, University of London, Egham, TW20 0EX, United Kingdom, [3]Chalmers University of Technology, Göteborg, S-412 96, Sweden, [4]Istituto Nazionale di Ricerca Metrologica, Strada delle Cacce, 91-10135 Torino, Italy, [5]Linköping University, Linköping, S-581 83, Sweden.

Using local scanning electrical techniques we study edge effects in side-gated Hall bar nanodevices made of epitaxial graphene. We demonstrate that lithographically defined edges of the graphene channel exhibit hole conduction within the narrow band of ~60–125 nm width, whereas the bulk of the material is electron doped. The effect is the most pronounced when the influence of atmospheric contamination is minimal. We also show that the electronic properties at the edges can be precisely tuned from hole to electron conduction by using moderate strength electrical fields created by side-gates. However, the central part of the channel remains relatively unaffected by the side-gates and retains the bulk properties of graphene.

Graphene is a two-dimensional material comprising carbon atoms closely packed in a honeycomb crystal structure[1]. Local layer thickness, substrate and environmental doping influence the material's nanoscale electronic properties. Due to the two-dimensional nature of graphene scanning probes can easily access its electronic states[2,3]; on the other hand the properties of the material are easily affected by the surrounding environment.

Electrical gating of graphene devices is a powerful technique providing an additional degree of freedom in manipulation of the carriers and allowing for a precise control of electronic nanodevices. For example, the carrier density and even type of carriers in exfoliated graphene have been extensively controlled from p- to n-type using semiconducting substrates with an insulating top layer, i.e. Si/SiO$_2$, allowing for back-gating[1,4–6]. Top-gates can also be used to achieve the same result[7–9]. However, they typically require deposition of a dielectric, such as HfO$_2$, on top of the graphene, which makes fabrication process more complicated, can degrade the carrier mobility[10] and also makes the surface of graphene inaccessible for sensing applications. In systems such as epitaxial graphene on semi-insulating SiC, side-gates could provide a valuable alternative as they allow for the doping control, while retaining accessibility to the graphene surface. For example, Chen *et al.* demonstrated that the doping can be modulated with side-gated sub-micron scale devices fabricated out of single layer exfoliated graphene deposited on highly doped Si substrate[11].

Epitaxially grown graphene on SiC has shown great promise for commercialisation due to wafer-scale production[9]. However, back-gating has proved difficult due to the semi-insulating nature of the substrate[12]. Research into nanometre scale side-gated devices fabricated out of epitaxial graphene on Si-face of 4*H*-SiC has shown a high transconductance[13], whereas field-effect transistors fabricated from multilayer graphene grown on C-face of SiC have a relatively low conductance[7]. These results show that side-gates can potentially provide an effective way of controlling the electronic properties of graphene[14].

Recently, the edges of graphene nanoribbons (GNR) have attracted much interest due to their strong influence on electronic and magnetic properties. The direct proof of the distinctive edge states in GNR and graphene quantum dots characterised by altered electronic properties was obtained by scanning tunnelling microscopy and scanning tunnelling spectroscopy techniques for different types of graphene[15–18]. At the same time, the influence of edges on electronic properties of 'bulk' graphene (including submicron graphene channels in lithographically defined devices) has received relatively little attention. For example, indirect proof of the edge states characterised by high density of defects and increased doping level was demonstrated by Raman microscopy[19]. Additional Raman microscopy studies demonstrated predominant p-doping in a graphene antidot lattice, which was attributed to the effect of edge states[20]. Another indirect example was presented by Chae *et al.*, where a strong enhancement of the edge conduction was explained in terms of confined charge inversion at the edge of an exfoliated graphene sheet[21]. Finally, the most direct proof of charge inversion at the channel edges of exfoliated graphene in vicinity of the Dirac point was demonstrated by scanning photocurrent microscopy[22]. Understanding the edge states is essential for measurements in the quantum Hall regime[23], edge photocurrents[24] and GNR devices[25], where the electronic properties are dominated by edge effects. In addition, it has been proposed that defective edges can also be used for Li storage applications[26].

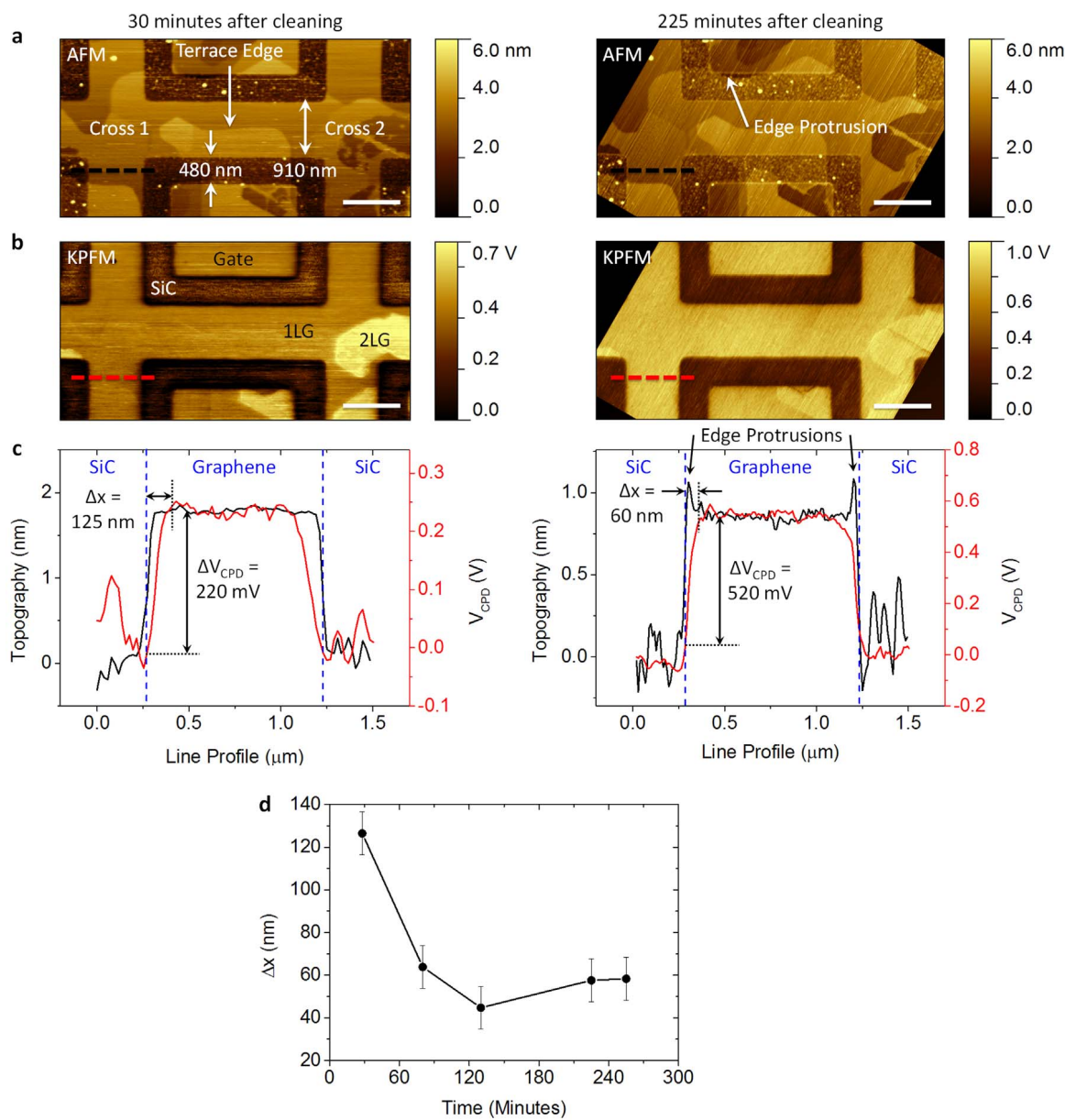

**Figure 1 | Topography and surface potential maps of epitaxial graphene device.** (a) Topography and (b) $V_{CPD}$ maps of device #1 taken 30 and 225 minutes after the contact-mode AFM cleaning process. White scale bars in (a) and (b) are 1 μm. (c) Topography and $V_{CPD}$ line profiles across the black and red dashed lines indicated in (a) and (b), respectively. On the topography profile taken 225 minutes after the cleaning process, small protrusions appear at the graphene edges, which are associated with preferential decoration by adsorbent molecules. Blue vertical dashed lines in (c) indicate geometrical edges of the graphene. The surface potential measurements were obtained with all electrical contacts grounded, i.e. $V_g = 0$. (d) Time dependence of the edge region width (Δx).

The electronic properties of graphene devices are typically studied using transport measurements, which require samples to be lithographically patterned and electrically contacted to form a Hall bar device. Transport measurements only shed light on the bulk electronic properties of the device and does not allow for direct studies of the edge effects. On the other hand, functional electrostatic force microscopy techniques provide local information about surface potential ($V_{CPD}$), work function (Φ), carrier density (n), resistance (R), and carrier mobility (μ) on the nanoscale[27–30].

In this work, we use a combination of standard transport measurements with scanning probe microscopy (SPM) modes, such as Kelvin probe force microscopy (KPFM)[31–34] and electrostatic force spectroscopy (EFS)[35], to study inversion of the carrier type in side-gated micron-scale epitaxial graphene devices. We show that while bulk of the epitaxial graphene is electron doped ($n_e \sim 2.95 \times 10^{12}$ cm$^{-2}$), the lithographically defined edges exhibit intrinsic hole conduction ($n_h \sim 7.33 \times 10^{11}$ cm$^{-2}$) within the width of ~60 nm, in the absence of side-gating ($V_g = 0$). The carrier inversion effect at the edges is the most pronounced immediately after the contact-mode atomic force microscopy (AFM) cleaning process, i.e. when the carrier distribution is affected primarily by the substrate doping. Furthermore, we demonstrate that the electronic properties at the edges can be precisely tuned from hole to electron conduction by using electrical side-gates. These results are also supported by electrostatic simula-tions of the graphene device in the presence of side-gates with variable voltage.

## Results

**Morphology and surface potential mapping of graphene nanodevices.** Topography height image of device #1, obtained using tapping mode AFM 30 minutes after the cleaning process, shows that the sample is essentially clean from resist residues (Figure 1a, c). However, topography height image, acquired 225 minutes after the cleaning process, shows small protrusions (~0.2 nm height) at the edges of the graphene (Figure 1a, c). These protrusions are likely to be attributed to adsorbent molecules from the ambient air (temperature = 22°C, ~40% relative humidity), that are attracted to chemically active defective states at the graphene edge. These adsorbates can be radical groups containing hydrocarbons, nitrogen and/or oxygen[36] and, therefore, providing soft gating to graphene at the edges. Furthermore, the effect from the adsorbates can be reversed by vacuum treatment and/or by heating. Overall, the topography is strongly dominated by wave-like steps associated with terrace edges (~0.25–1 nm height) in the SiC substrate (Figure 1a).

The $V_{CPD}$ map of the grounded device, obtained using frequency-modulated (FM)-KPFM, reveals predominantly single-layer graphene (1LG) with a few small patches of bi-layer graphene (2LG) (Figure 1b). The differences in the absolute values of the $V_{CPD}$ are the result of differences in the work functions of the probe, 1LG, 2LG and SiC. The $V_{CPD}$ line profiles, obtained 30 and 225 minutes after the cleaning process, reveal a contrast of ~220 and ~520 mV, respectively, between the graphene channel and etched SiC (Figure 1c). The graphene-SiC edge was determined from the topography line profiles (Figure 1c) by using the edge spread function, as defined in the Standard on Lateral Resolution (ISO/TR 19319:2013. Surface chemical analysis -- Fundamental approaches to determination of lateral resolution and sharpness in beam-based methods.), by taking the half-distance between two points of well-defined relative intensity (12 and 88%). The $V_{CPD}$ value indicates that initially the electrical properties of graphene up to ~125 nm from the edge are inherently different from bulk (Figure 1c and 1d), presumably due to crystalline defects. However, a few hours after the cleaning process, the affected region decreases to ~60 nm (Figure 1c and 1d), due to appearance of edge protrusions observed during AFM imaging (Figure 1a and 1c) and respective environmental doping.

**Nanoscale visualisation of the side-gating effect.** The effect of the side-gates was investigated on device #1 by measuring the $V_{CPD}$ between the probe and gate-SiC-channel-SiC-gate, at $V_g = 0$, $\pm 1$ and $\pm 2$ V with FM-KPFM (Supplementary Information, Figure S1a). The vertical dashed lines indicate the graphene-SiC boundaries for the channel and gates. The results show a negligible change in the $V_{CPD}$ across the central part of the channel, however, the close look at the edges shows a clear dependence on $V_g$ (Figure 2a). To differentiate the effect of the side-gates from the inherent changes in the work function at the edge of the channel, the $V_{CPD}$ was normalised using the grounded line profile, $\Delta V_{CPD} = V_{CPD}(V_g) - V_{CPD}(0)$ (Supplementary Information, Figure S1b). In the case of $V_g = +2$ V, $\Delta V_{CPD}$ increases by ~80 mV at the lithographically defined edges of channel, whereas for $V_g = -2$ V, $\Delta V_{CPD}$ decreases by ~270 mV (Figure 2b). These results indicate that negative side-gate voltages are over 3 times more effective in changing the $V_{CPD}$ at the edges of the channel.

**Nanoscale quantification of the side-gating effect.** Using EFS with a calibrated Pt-Ir probe, the work functions of 1LG at the centre ($\Phi_C$) and ~30 nm from the edge ($\Phi_E$) at $V_g = 0$ were determined to be $\Phi_C = 4.17 \pm 0.06$ eV and $\Phi_E = 4.47 \pm 0.08$ eV, respectively (Figure 3a). $\Phi_C$ is somewhat lower than previously published values, i.e. 4.4–4.6 eV for n-type 1LG on 4H-SiC[30,37], which could be attributed to differences in the substrate and graphene growth parameters leading to a higher carrier concentration. This results in $\Delta \Phi_{C-E} = 300$ meV, indicating that the carrier density at the edges of the channel is significantly different to the bulk. To determine the carrier density at the edge of the channel:

- First, we define the position of the Dirac point as $E_D = \Phi_C + E_F = 4.37 \pm 0.06$ eV (Figure 3b), using the work function at the centre ($\Phi_C = 4.17$ eV) and Fermi energy obtained from bulk transport measurement ($E_F = 200$ meV).
- Second, we define the Fermi energy at the edge as $E_F = E_D - \Phi_E = -100$ meV (Figure 3b) using $E_D$ and work function at the edge ($\Phi_E = 4.47$ eV). In this case, the negative sign of $E_F$ indicates p-type conduction at the edge.
- Finally, the inherent carrier density at the edge (at $V_g = 0$) was calculated as $n_h = 7.33 \times 10^{11}$ cm$^{-2}$ using the Fermi energy at the edge ($E_F = -100$ meV) and $n_h = \frac{1}{\pi}\left(\frac{E_F}{\hbar v_F}\right)^2$.

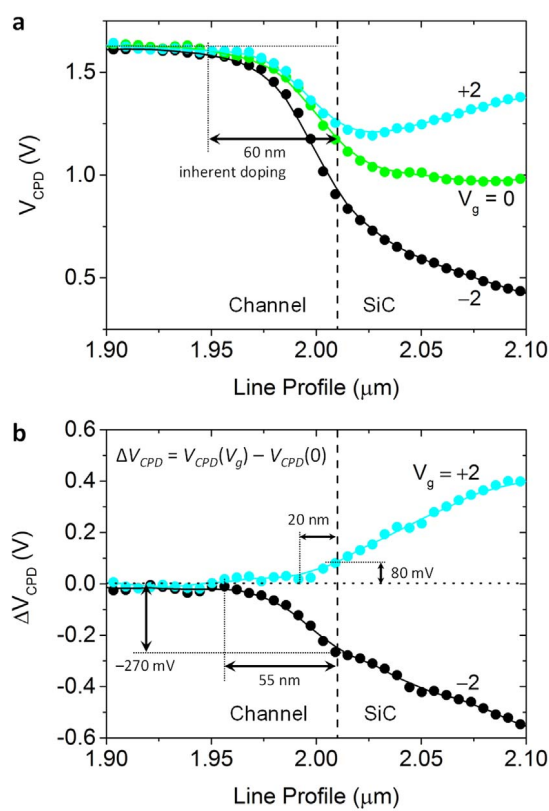

**Figure 2 | Visualisation of the edge effects.** (a) $V_{CPD}$ and (b) $\Delta V_{CPD} = V_{CPD}(V_g) - V_{CPD}(0)$ line profiles at the edge of the graphene channel. The vertical dashed lines mark out the graphene-SiC edge. The line profile at $V_g = 0$ in (a) shows that electronic properties up to ~60 nm from the edge of the channel are inherently different compared to bulk. The normalised line profiles in (b) take into account the inherent changes in the work function of the device, thereby only showing $\Delta V_{CPD}$ as a result of the side-gates. Solid lines are a guide for the eye only.

While n-type conduction dominates the central region of the device, the transition to p-type conduction at the edge has also been previously revealed using scanning gate microscopy[21] and illumination with circularly polarised terahertz radiation[24]. Inversion of the carrier type at the edge is attributed to the defective crystalline structure and adsorbed molecules attached to the dangling bonds and acting as local dopants.

The effectiveness of the side-gates was accurately quantified by performing a series of EFS measurements at the centre and ~30 nm from the edge of the channel on device #2 (Figure 3a inset), using the same calibrated Pt-Ir probe. The 280-nm channel-gate spacing results in electric field strength averaged over the channel edge of $E \sim 160$ kVcm$^{-1}$ at $V_g = \pm 2$ V, see Supplementary Information 2.1). Over 400 EFS measurements were performed at each point by sweeping the applied gate voltage from $V_g = -2$ to $+2$ V in increments of 100 mV, giving an average of 10 measurements at each $V_g$. The raw EFS data was processed to extract the $V_{CPD}$ and, thus, the work function at each $V_g$ (Figure 3c). At the centre of the channel (black squares) the overall $V_g$-induced change in the work function was $\pm 90$ meV. On the other hand, at the edge (red circles) the work function was significantly modified and values of $\Delta\Phi_E = \Phi_E(V_g) - \Phi_E(0) \sim 345 \pm 75$ meV and $\Delta\Phi_E \sim 167 \pm 110$ meV were measured at $V_g = -2$ and $+2$ V, respectively, showing a total change of ~512 meV. Figure 3d shows the carrier density at the centre (black squares) and edge (red circles) of the channel, which was determined by calculating $E_F$ from the work function measurements (Figure 3c) and the previously calculated value of $E_D = 4.37$ eV using $\pm E_F = E_D - \Phi_{E,C}$, where $+/-$ indicates n-/p-type, respectively. Figure 3d shows that $n_e$ at the centre of the channel is largely unaffected by $V_g$, remaining n-type. However, the p-type conduction observed at the edge exhibits a strong dependence on $V_g$, e.g. $n_h \sim 1.8 \times 10^{13}$ cm$^{-2}$ at $V_g = -2$ V, whereas the Dirac point was reached with the application of $V_g \sim +1$ V. Moreover, conductivity of the edges can be switched to n-type with $n_e \sim 1.3 \times 10^{12}$ cm$^{-2}$ by applying $V_g = +2$ V. One might also consider the probe to induce an additional gating effect. However, the measurement is performed when the probe-sample potential difference is zero, thus eliminating any probe-sample capacitive coupling. The carrier doping effects due to side-gates were also theoretically investigated by deducing the spatial variation of the induced surface carrier density and modification of the edge work function from electrostatic field simulations, performed with a 3D boundary element based code that includes the presence of SiC substrate under the graphene sheet (Supplementary Information, Figures S3, S4 and S5). It was demonstrated that the electrical gating considerably affects carrier concentration up to a distance of some tens of nanometres from the interface with SiC and the induced charge varies linearly with $V_g$.

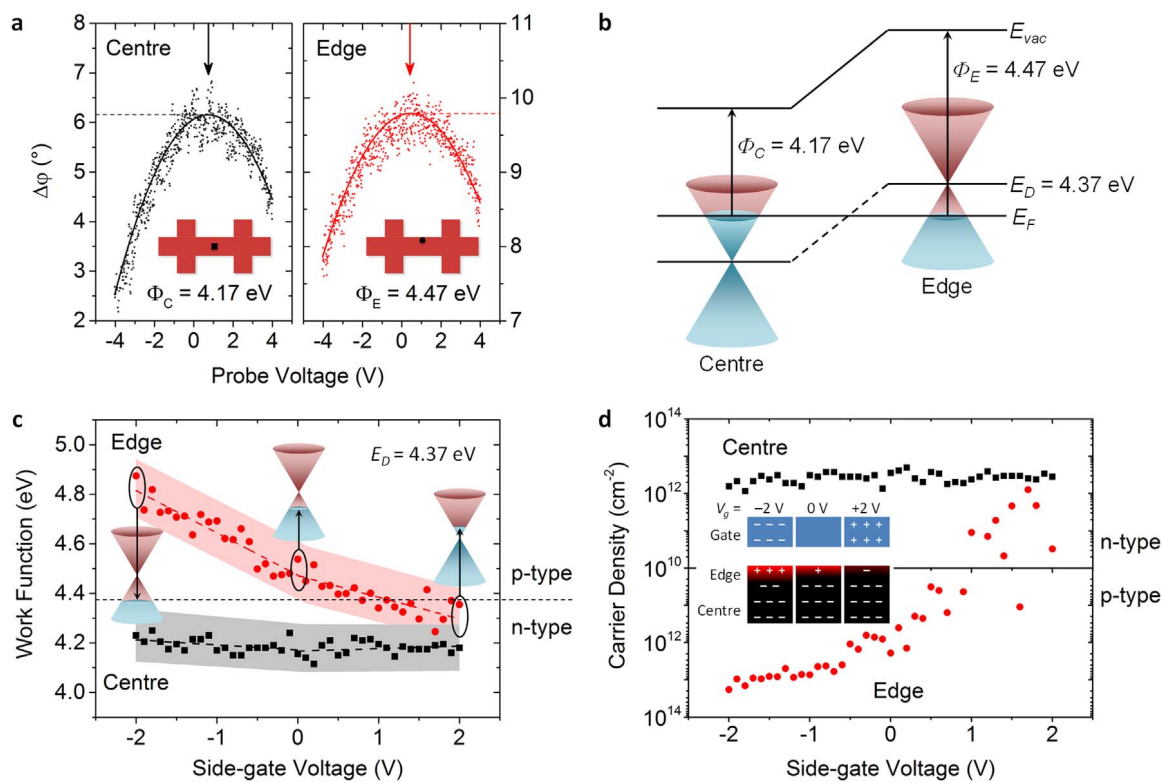

**Figure 3 | Quantification of the edge effects.** (a) EFS measurements performed on device #2 at the centre and ~30 nm away from the edge of the graphene channel as indicated by the respective schematics. Solid lines show the averaged values of the phase change. (b) Schematics of the band diagram for 1LG showing the Fermi energy and work function at the centre and edge of graphene channel at $V_g = 0$. Dependence of the (c) work function and (d) carrier density on the side-gate voltage at the centre (black squares) and ~30 nm away from the edge (red circles) of the channel. The work function was measured with EFS for a grounded channel, while sweeping $V_g$. The inset in (c) shows a schematic of the band diagram for 1LG at the edge of the channel at different $V_g$. The inset in (d) shows a schematic of the charge distribution at the centre and edge of the channel. The dashed lines in (c) are a guide for the eye only and the shaded areas represent the error bars.

If we consider the case of $V_g > 0$, electrons are attracted towards the edge of the channel (Figure 3d inset). By definition, the work function is the minimum energy required to remove an electron from the Fermi energy into vacuum; thus, a reduction in the work function at $V_g > 0$ is due to an increase in the electron concentration at the edge. The opposite scenario occurs when $V_g < 0$.

**Effect of side-gating on bulk electronic properties.** The effect of the side-gating on the *bulk* transport properties of devices #2, #3 and #4 was investigated by measuring the carrier density, $n_e$, and the 4-point resistance variation, $\Delta R_4 = R_4(V_g) - R_4(0)$, (Figures 4a and 4b, respectively). The base carrier density ($n_e \sim 2.95 \times 10^{12}$ cm$^{-2}$) remained completely unaffected by the side-gates at these relatively moderate electric fields, as the measurements were conducted in the Hall cross geometry, where the Hall voltage is significantly less influenced by $V_g$ due to the geometry of the experiment. Resistance measurements were performed along the channel surrounded by the gates. In this case sweeping $V_g$ from $-5$ to $+5$ V led to a measurable change in the channel resistance by a total of up to $\pm 15 \,\Omega$ on top of $R_4 = 8.9$ k$\Omega$ for device #2, which corresponds to a total change in resistance $\Delta R_4 \sim 0.33\%$. Resistance measurements conducted on devices #3 and #4 with narrower channels and the same width of etched SiC trenches showed that $\Delta R_4$ is inversely proportional to the

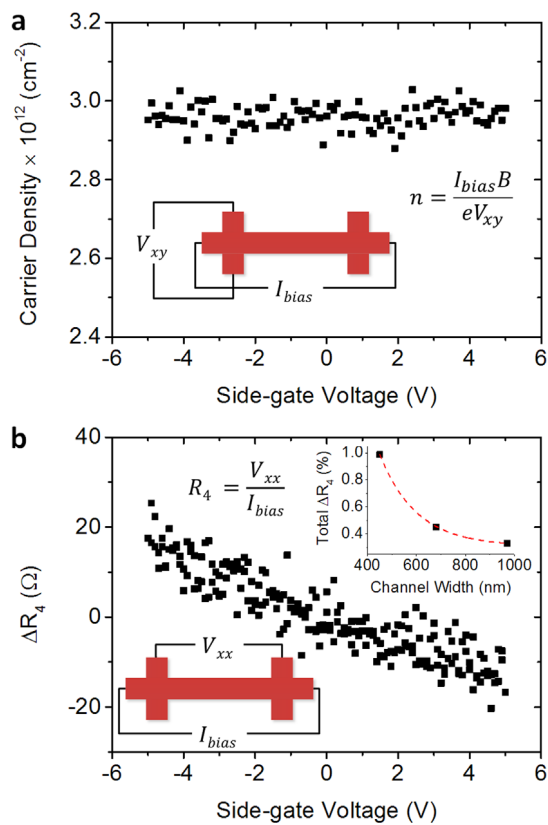

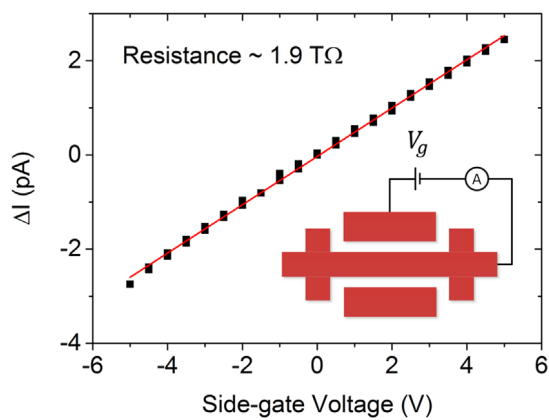

Figure 5 | **Leakage current.** Bulk transport measurements showing dependence of the normalised leakage current on the gate voltage, $\Delta I = I(V_g) - I(0)$, in device #2. The Ohmic fit (solid line) reveals the gate-channel resistance of $\sim 1.9$ T$\Omega$. Inset shows the schematics of the measurement, which was performed between the channel and the side-gate.

**Summary.** In summary, we performed local work function and bulk transport studies to evaluate the edge states in graphene devices caused by geometric boundaries. We demonstrate the direct observation of edge carriers' inversion in epitaxial graphene even in the absence of a back gate, i.e. significantly far from the Dirac point and at relatively high carrier concentration. We showed that, while the bulk of the material is n-doped, the edges of graphene devices are inherently p-doped. One of the possible reasons is the plasma etching process used to define the device, which might introduce defects both in the graphene and in the underlying interfacial layer, affecting the electronic properties of the graphene edges. The carrier inversion effect at the edges is the most pronounced immediately after the cleaning process. We also showed that over the course of a few hours after the cleaning, protrusions appear at the lithographically defined edges. These are likely to be adsorbates attracted by the chemically active defective states, thus providing additional soft gating to graphene at the edges.

We further studied the effect of the electrical gating in the side-gate geometry on the surface potential in graphene devices, together with electrostatic modelling, showing that the electronic properties can be influenced up to tens of nanometres from the device edge (in the given geometry and state of the device and field strength). No modification of the work function was observed at the centre, however a clear change of $\Delta\Phi_E \sim 512$ meV was measured at the edge of the channel at $V_g = \pm 2$ V. In this case, the near-edge region could be precisely tuned from hole to electron conduction with $n_h \sim 1.8 \times 10^{13}$ cm$^{-2}$ to $n_e \sim 1.3 \times 10^{12}$ cm$^{-2}$, respectively. Controlled transport measurements revealed that the side-gate voltage is only responsible for a very small change in the total resistance ($\sim 1\%$) of a 450 nm-wide channel. However, the edge effects are certain to play an increasing role in smaller devices, paving the way for nanoscale electrically controlled epitaxial graphene devices.

Thus, local electrical measurements allow for linking of electrical properties to the device geometry and defective states and can be used to understand the characteristics of the whole device as defined through bulk transport measurements. Understanding the edge effects in graphene is essential for quantum Hall regime applications, edge photocurrents and graphene nanoribbons devices.

## Methods

**Device fabrication.** Nominally monolayer epitaxial graphene was prepared by sublimation of SiC and subsequent graphene formation on the Si-terminated face of semi-insulating on-axis 4H-SiC(0001) substrate at 2000°C and 1 bar argon gas pressure. Details of the growth and structural characterisation are reported elsewhere[38].

Figure 4 | **Side-gating effect on the bulk transport properties.** Bulk transport measurements showing dependence of (a) carrier density and (b) change in the 4-terminal resistance, $\Delta R_4 = R_4(V_g) - R_4(0)$, on the side-gate voltage in device #2. Top inset in (b) shows the total $\Delta R_4$ at $E \sim \pm 400$ kVcm$^{-1}$ for devices #2, #3 and #4, where the dashed line is a guide for the eye only. Measurement schematics for $n$ and $R_4$ are shown in insets (a) and (b), respectively.

channel width with a maximal value of $\Delta R_4 \sim 1\%$ at $w = 450$ nm (Figure 4b inset). These measurements show that side-gating is more effective for narrower devices. Thus, transport measurements imply that, although the bulk of the material is relatively unaffected by the side-gates, the edges accounting for a significant portion of the total channel width, play a significant role in smaller devices. The variation of the device resistance with gate voltage was also theoretically investigated by means of a two-fluid based transport model of the graphene sheet, under the assumptions of diffusive transport regime and stationary conditions (Supplementary Information, Figure S7). In particular, side-gate effects were included via a spatially dependent carrier density, which is a function of gate voltage.

**Gate-channel leakage current.** Further, we study whether the current leaking through the SiC substrate could affect the electronic properties of graphene, as, for instance, the case in Ref. 7. Figure 5 shows the dependence of the normalised leakage current, $\Delta I = I(V_g) - I(0)$, on $V_g$. The normalisation against $I(V_g = 0)$ is essential for each $I(V_g)$ measurement for precise determination of the low current due to charging of the insulating substrate, which leads to hysteretic behaviour in the leakage current. Using this method, we are able to determine that the resistance between the channel and the gates is $\sim 1.9$ T$\Omega$ and linear in the range $+/- 5$ V. Thus, in the side-gate experiments, the maximum current leaking from the gate to the channel is $\sim 2$ pA at $V_g = \pm 5$ V, which is one part in a million of the $I_{bias}$ used for the transport measurements. Thus, we can conclude that the side-gate effects observed in this work are the result of the electric field only and not caused by the leakage current.

| Table 1 | Geometrical parameters of the devices | |
|---|---|---|
| device | channel width (nm) | gate-spacing (nm) |
| #1 | 910 | 480 |
| #2 | 970 | 280 |
| #3 | 680 | 280 |
| #4 | 450 | 280 |

The double-cross Hall devices were fabricated from epitaxial graphene by three steps of e-beam lithography (EBL), oxygen plasma etching and evaporation of Cr/Au (5/100 nm) electrodes for contacting[39,40]. The resulting material is n-doped, owing to charge transfer from the interfacial layer[39,41]. The bulk electron concentration in ambient conditions was determined by sweeping the out-of-plane magnetic field up to B = 0.5 T and measuring the Hall voltage ($V_H$) at applied bias current ($I_{bias}$)[39,42], $n_e = \frac{I_{bias}B}{eV_H} = 2.95 \times 10^{12}$ cm$^{-2}$. The carrier mobility was defined as $\mu = \frac{I_{bias}}{neV_{xx}} \times \frac{L}{W}$ = 1270 cm$^2$V$^{-1}$s$^{-1}$, where $e$ is the electronic charge, $V_{xx}$ is the longitudinal voltage measured from crosses 1 and 2, $L$ is the distance between the centres of the two crosses and $W$ is the width of the channel. The mean free path, $\lambda = \frac{h}{2e}\mu\left(\frac{n}{\pi}\right)^{1/2} \sim 26$ nm, and the Fermi energy, $E_F = \hbar v_F \sqrt{\pi n}$ = 200 meV, were also calculated.

Four different devices of similar layout were measured. Device #1 consists of two symmetric crosses with $w$ = 910 nm-width channel, surrounded by a trench with the width $d$ = 480 nm etched down into the SiC substrate (Figure 1a). Devices #2, #3 and #4 consist of $w$ = 970, 680 and 450 nm-wide channels, respectively, with $d$ = 280 nm (Table 1). The isolated areas of graphene next to the channel were used as side-gates. Relatively moderate electric fields (up to an average value over the channel edge of $E$ ~ 400 kVcm$^{-1}$ at $V_g = \pm 5$ V) were used, being generally limited by the 3.23 eV bandgap of 4H-SiC[43].

**Theoretical models.** The electric field and the carrier doping effects induced by side-gates are computed by means of a 3D boundary element code. The consequent variation of the device resistance is investigated by means of classical transport modelling. The models are described in the Supplementary Information Section 2.

**Sample cleaning.** It is well known that the EBL process tends to leave 1–2 nm thick resist residues on graphene devices[30,44]. This thin residue film is notoriously resistive to chemical cleaning and has been known to alter the carrier concentration of graphene. In the case of exposing the residues to large doses of deep UV or electron beam irradiation, the doping can be varied from n-type to p-type[30,45]. The residues can also attach to SPM probes, changing both their shape and work function. To avoid this contamination and study the properties of pristine graphene, contact-mode AFM was performed on a Bruker Icon SPM, using soft cantilevers (Bruker) with a set point of ~40 nN. The residues were mechanically scraped to the sides of the area of interest by scanning it several times until the desired cleanliness was achieved[44]. This procedure does not damage the graphene in anyway, as confirmed by transport measurements described in Ref. 30.

**SPM measurement.** The effect of the side-gates on electronic properties on the nanometre scale was investigated using SCM-PIT Pt-Ir coated probes (Bruker) with a probe radius of ~20 nm and a force constant of ~0.8 Nm$^{-1}$. $V_{CPD}$ map was obtained simultaneously with topography using FM-KPFM. FM-KPFM (EFS) operates on the electrostatic force gradient via frequency (phase) shift[30], which is most sensitive at the probe apex, resulting in spatial resolution of ~20 nm (Supplementary Information, Figure S8)[46]. For the comprehensive review of the experimental methods applied to graphene samples and devices similar to those described here, see Refs. 30 and 44. The changes to the electrical properties were further quantitatively studied using a point spectroscopy technique (EFS)[30,35]. Ten sets of EFS measurements were performed at each $V_g$ and each of the two well-defined points, i.e. at the centre and ~30 nm away from the edge of the channel of device #2 (Figure 3a inset), by sweeping $V_{probe}$ and measuring the phase change of the cantilever (Figure 3a). The resulting data has a parabolic response due to the electrostatic attractive/repulsive probe-sample interactions. When $V_{CPD} = V_{probe}$, the probe experiences no electrostatic forces, i.e. $d\varphi/dV_{probe} = 0$. The raw data was post-processed to accurately fit each parabola and extract the value of $V_{probe}$ at the inflection point, effectively providing a measurement of $V_{CPD}$. While scanning techniques such as FM-KPFM also provides a measure of the $V_{CPD}$, a significant issue arises when measurements are conducted on contaminated samples, which can result in the work function of the probe ($\Phi_{probe}$) changing due to pickup of foreign material and wear of the coating layer. However, changes to $\Phi_{probe}$ are largely negated by implementing the spectroscopy technique, where the precisely calibrated $\Phi_{probe}$ does not change throughout the measurement as no scanning is involved. The work function of the Pt-Ir probe ($\Phi_{probe}$ ~ 5.29 eV) was initially calibrated against the gold leads using the approximation: $\Phi_{probe} \sim \Phi_{Au} + eV_{CPD}$[30,33], where $\Phi_{Au}$ = 4.82 eV as obtained from ultraviolet photoelectron spectroscopy (UPS) measurements[44]. Although $\Phi_{Au}$ was determined in ultra-high vacuum with UPS measurements, we show that $\Phi_{Au}$ increases by <1% when transferred from high-vacuum to ambient humidity ~40% (Supplementary Information, Section 4 and Figure S9), making it ideal for calibrating $\Phi_{probe}$.

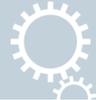

## Acknowledgments

We are very grateful to Sergey Kubatkin for directing us towards side-gated devices. We are thankful to T.J.B.M. Janssen and T.L. Burnett for useful discussions. This work has been funded by NMS under the IRD Graphene Project (NPL), EMRP under projects MetMags, GraphOhm and Graphene Flagship.

## Author contributions

O.K. and A.T. designed the research, R.Y. grew the epitaxial graphene, A.L. fabricated the devices, V.P. performed the experiments, V.P., A.T. and O.K. analysed the data and A.M. wrote the code and ran the model. All authors discussed the results. V.P., A.M., A.T. and O.K. participated in writing the manuscript and V.P., A.M. and O.K. participated in writing the Supplementary Information. All authors reviewed and commented on the manuscript.

## Additional information

**Supplementary information** accompanies this paper at http://www.nature.com/scientificreports

**Competing financial interests**: The authors declare no competing financial interests.

**How to cite this article:** Panchal, V. *et al*. Visualisation of edge effects in side-gated graphene nanodevices. *Sci. Rep.* **4**, 5881; DOI:10.1038/srep05881 (2014).

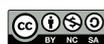